\documentclass[pre,superscriptaddress,twocolumn,a4paper,showpacs,nofootinbib]{revtex4-1}
\usepackage{graphics}
\usepackage{amssymb}
\usepackage{amsmath}
\usepackage{wasysym}
\usepackage{latexsym}
\usepackage{graphicx}
\usepackage{epsfig}
\usepackage{subfigure}
\usepackage{float}
\usepackage{color}
\usepackage{lineno}
\usepackage{verbatim}
\usepackage{enumitem}
\usepackage[T1]{fontenc}

\begin{document}

\title{Imitation dynamics and the replicator equation}

\author{Jos\'e F.  Fontanari}
\affiliation{Instituto de F\'{\i}sica de S\~ao Carlos,
  Universidade de S\~ao Paulo,
  Caixa Postal 369, 13560-970 S\~ao Carlos, S\~ao Paulo, Brazil}


\begin{abstract}
Evolutionary game theory has impacted many fields of research by providing a mathematical framework for studying the evolution and maintenance of social and moral behaviors.   This success is owed in large part to the demonstration that the central equation of this theory - the replicator equation - is the deterministic limit of a stochastic imitation (social learning)  dynamics.    Here we offer an alternative elementary proof of this result, which holds for the scenario where players compare their instantaneous (not average) payoffs to decide whether to maintain or change their strategies, and only more successful individuals can be imitated.
\end{abstract}

\maketitle

\section{Introduction}

From a selfish perspective, people only behave in ways that benefit themselves.  As social animals, acceptance in their social milieu is a crucial aspect of human  well-being, and so imitating better-off peers is likely to be a major behavioral drive \cite{Bandura_1977,Blackmore_2000}. The analysis of this scenario is particularly well suited to the evolutionary game-theoretic framework originally introduced to model animal contests \cite{Maynard_1973,Maynard_1982}. In fact, evolutionary game theory has been used extensively to address a problem that has been heralded as one of the greatest challenges for science in the twenty-first century \cite{Kennedy_2005}: understanding and promoting cooperation in human societies within a scenario of selfish individuals (see, e.g., \cite{Axelrod_1984,Sigmund_2010,Pacheco_2014,Perc_2017,Wang_2023,Fontanari_2024}).

The use of evolutionary game theory to study social, as well as moral \cite{Capraro _2018,Capraro _2019,Vieira_2024}, behaviors has been greatly stimulated by the proof that the central equation of this theory -- the replicator equation \cite{Hofbauer_1998,Nowak_2006} -- describes the dynamics of imitation (or social learning) in the limit of infinite population size \cite{Traulsen_2005} (see also \cite{Sandholm_2010}).  Because of the importance of this result in so many areas of research, we feel it necessary to give an elementary proof.  More importantly, we give a proof that holds for the implementations of the stochastic imitation dynamics where players compare their instantaneous (not average) payoffs and only more successful individuals can be imitated.

\section{Imitation dynamics}
Consider a well-mixed finite population of size $M$, consisting of individuals that can play either strategy $A$ or strategy $B$.  The population is well-mixed in the sense that every individual can interact with every other individual in the population.
At each time step $\delta t$, a focal individual $i$ and a model individual $j$ are randomly selected from the population without replacement. 
If the focal individual's instantaneous payoff $f_i$ is less than the model individual's instantaneous payoff $f_j$,  the focal individual copies the model's strategy with probability
\begin{equation}\label{prob0}
 \frac{f_{j} - f_{i}}{\Delta_m}, 
\end{equation}
where $\Delta_m$ is the maximum possible payoff difference that guarantees that this ratio is less than or equal to 1.  If $f_{j} \leq f_{i}$, then the focal individual maintains its strategy.  We emphasize that the instantaneous payoffs $f_i$ and $f_j$ are very unlikely to be the result of a game between the focal and model individuals, as we will see next when discussing the $2$-player and $N$-player games.
As usual in such an asynchronous update scheme, we choose the time step $\delta t = 1/M$ so that during the step from $t$ to $t+1$ exactly $M$, though not necessarily different, individuals are selected as focal individuals. 

Equation (\ref{prob0}) differs in two significant ways from the switching probability used in the proof offered by Ref.\ \cite{Traulsen_2005} for the convergence of the imitation dynamics to the replicator equation as the population size goes to infinity. First and foremost, in Ref.\ \cite{Traulsen_2005}, instantaneous payoffs are replaced by average payoffs, which means that the focal and model individuals play many games against all members of the population and accumulate payoffs before engaging in the imitation process. Second, in Ref.\ \cite{Traulsen_2005}, the probability (\ref{prob0}) appears as a perturbation to a baseline payoff-independent switching probability and is also valid when $f_i > f_j$, so there is a positive probability that individuals would imitate the strategy of a peer with a lower average payoff.
Both of these assumptions are too restrictive and unnecessary. 

In fact, imitation of more (not less) successful individuals is a key feature of social learning theory \cite{Bandura_1977} and has been hailed as the fabric of human society \cite{Blackmore_2000}, in addition to being the inspiration for a variety of effective optimization algorithms \cite{Kennedy_1998,Fontanari_2014}.
Of course, since uncertainty is ubiquitous in real life, a more realistic imitation scenario requires relaxing the condition that only
more successful individuals can be imitated. Indeed, there is evidence that the  imitation process (social learning) in problem solving experiments is best described by introducing a noise factor into the imitation rule \cite{Toyokawa_2019}.  A popular imitation rule that incorporates a noise parameter is the Fermi strategy update rule \cite{Szabo_1998,Perc_2010}, where the temperature regulates the relevance of payoffs on the focal individual's decision to change or not to change strategy. Note, however, that if the imitation rule is a nonlinear function of the payoff difference $f_{j} - f_{i}$, as is the case for the Fermi rule, then the resulting imitation dynamics is not described by the replicator equation in the infinite population limit, as we will show here.

Furthermore, there are games, such as the $N$-player majority vote \cite{Soares_2024}, for which the mean, but not the instantaneous, payoffs of individuals are identical. In this case, using the mean payoffs in the switching probability would freeze the stochastic dynamics in the initial conditions.  The implementation of the stochastic imitation dynamics using the switching probability (\ref{prob0}) is common in the evolutionary game literature (see, e.g., \cite{Zheng_2007,Meloni_2009}) but  a simple and accessible proof linking this dynamics to the replicator equation seems to be lacking.

Here we provide such a proof. For didactic purposes, we consider separately $2$-player games, where the payoffs are determined by pairwise interactions, and $N$-player games, where the payoffs cannot be reduced to pairwise interactions \cite{Perc_2013}.

\section{$2$-player games}

Classical examples of $2$-player games are the hawk-dove game, used to introduce the concept of evolutionarily stable strategy \cite{Maynard_1982}, and the prisoner's dilemma, used to study the problem of cooperation in social dilemmas \cite{Axelrod_1984}.
The payoff  of a player using strategy $A$ or $B$ depends on the other player's strategy and is determined by the payoff matrix \cite{Maynard_1982}
\begin{equation}\label{mat}
\begin{array}{ccc}
 & A & B \\
A & a & b \\
B& c & d
\end{array} .
\end{equation}
More explicitly, individual $i$ chooses a random individual in the population and plays a game. This single game determines only the instantaneous payoff $f_i$ of individual $i$.  This procedure is done for the $M$ individuals in the population. Thus, at a given time $t$, the population is fully characterized by the number of individuals playing strategy $A$ with payoff $a$ ($I_a$), the number of individuals playing strategy $A$ with payoff $b$ ($I_b$), the number of individuals playing strategy $B$ with payoff $c$ ($J_c$), and the number of individuals playing strategy $B$ with payoff $d$ ($J_d$).  Of course, $I_a+I_b+J_c+J_d = M$. 

Let $P_i^a(t)$ and $P_i^b(t)$ be the probabilities that player $i$  uses strategy $A$ and has instantaneous payoff $a$ and $b$, respectively, at time $t$. Similarly, $Q_i^c(t)$ and $Q_i^d(t)$ are  the probabilities that player $i$ uses strategy $B$ and has instantaneous payoff $c$ and $d$, respectively, at time $t$. We will calculate  the conditional  probability  that player $i$  uses strategy $A$  at time $t + \delta t $  as a result of  the imitation dynamics. The probability is conditional on the state of the population at time $t$, which is described by the integers
$I_a$, $I_b$, $J_c$, and $J_d$.  We emphasize that the imitation dynamics only determines the players' strategies: their particular instantaneous payoffs are determined by the games described above.   Of course, the probability  that player $i$  uses strategy $A$  at time $t$ is $P_i(t) = P_i^a(t)+ P_i^b(t)$, and  the probability  that player $i$  uses strategy $B$  at time $t$ is $Q_i(t) = Q_i^c(t)+ Q_i^d(t)$ with
$P_i(t) + Q_i(t) = 1$.
The desired conditional probability, $P_{i}(t+\delta t)$,  is  given by the sum of the probabilities of the following exclusive events.
 \begin{enumerate}[label=(\alph*)]
\item Individual $i$ uses strategy  $A$ at time $t$ and another individual is chosen as the focal individual.  The probability of this event is 
\begin{equation}\label{item1}
\left [ P_{i}^a(t) + P_{i}^b(t) \right ]  \times \frac{M-1}{M}  .
\end{equation}
\item Individual $i$ uses strategy $A$ at time $t$ and is selected as the focal individual. The model individual also uses strategy $A$. The probability of this event is 
\begin{equation}\label{item2}
\left [ P_{i}^a(t) + P_{i}^b(t) \right ] \times \frac{1}{M} \times \frac{I_a+I_b-1}{M-1}.
\end{equation}
\item Individual $i$ uses strategy $A$ and has payoff $a$ at time $t$ and is selected as the focal individual. The model individual uses strategy $B$ and has payoff $c$, but individual $i$ maintains strategy $A$. The probability of this event is 
\begin{equation}\label{item3}
 P_i^a(t) \times \frac{1}{M} \times \frac{J_c}{M-1}  \left  [ 1   
- [ 1- \theta(a-c)]  \frac{c-a}{\Delta_m} \right ] .
\end{equation}
\item Individual $i$ uses strategy $A$ and has payoff $b$ at time $t$ and is selected as the focal individual. The model individual uses strategy $B$ and has payoff $c$, but individual $i$ maintains strategy $A$. The probability of this event is 
\begin{equation}\label{item4}
 P_i^b(t) \times \frac{1}{M} \times \frac{J_c}{M-1}  \left  [  1  
- [ 1- \theta(b-c)]  \frac{c-b}{\Delta_m}   \right ] .
\end{equation}
\item Individual $i$ uses strategy $A$ and has payoff $a$ at time $t$ and is selected as the focal individual. The model individual uses strategy $B$ and has payoff $d$, but individual $i$ maintains strategy $A$. The probability of this event is 
\begin{equation}\label{item5}
 P_i^a(t) \times \frac{1}{M} \times \frac{J_d}{M-1}  \left  [  1  
- [ 1- \theta(a-d)]  \frac{d-a}{\Delta_m}   \right ] .
\end{equation}
\item Individual $i$ uses strategy $A$ and has payoff $b$ at time $t$ and is selected as the focal individual. The model individual uses strategy $B$ and has payoff $d$, but individual $i$ maintains strategy $A$. The probability of this event is 
\begin{equation}\label{item6}
 P_i^b(t) \times \frac{1}{M} \times \frac{J_d}{M-1}  \left  [  1  
- [ 1- \theta(b-d)]  \frac{d-b}{\Delta_m}   \right ] .
\end{equation}
\item Individual $i$ uses strategy $B$ and has payoff $c$ at time $t$ and is selected as the focal individual. The model individual uses strategy $A$ and has payoff $a$, and individual $i$ changes to strategy $A$. The probability of this event is 
\begin{equation}\label{item7}
 Q_i^c(t) \times \frac{1}{M} \times \frac{I_a}{M-1}  \left  [  \theta(a-c)]  \frac{a-c}{\Delta_m}   \right ] .
\end{equation}
\item Individual $i$ uses strategy $B$ and has payoff $d$ at time $t$ and is selected as the focal individual. The model individual uses strategy $A$ and has payoff $a$, and individual $i$ changes to strategy $A$. The probability of this event is 
\begin{equation}\label{item8}
 Q_i^d(t) \times \frac{1}{M} \times \frac{I_a}{M-1}  \left  [  \theta(a-d)]  \frac{a-d}{\Delta_m}   \right ] .
\end{equation}
\item Individual $i$ uses strategy $B$ and has payoff $c$ at time $t$ and is selected as the focal individual. The model individual uses strategy $A$ and has payoff $b$, and individual $i$ changes to strategy $A$. The probability of this event is 
\begin{equation}\label{item9}
 Q_i^c(t) \times \frac{1}{M} \times \frac{I_b}{M-1}  \left  [  \theta(b-c)]  \frac{b-c}{\Delta_m}   \right ] .
\end{equation}
\item Individual $i$ uses strategy $B$ and has payoff $d$ at time $t$ and is selected as the focal individual. The model individual uses strategy $A$ and has payoff $b$, and individual $i$ changes to strategy $A$. The probability of this event is 
\begin{equation}\label{item10}
 Q_i^d(t) \times \frac{1}{M} \times \frac{I_b}{M-1}  \left  [  \theta(b-d)]  \frac{b-d}{\Delta_m}   \right ] .
\end{equation}
\end{enumerate}

Here we have used $\theta(x) = 1$ if $x\geq 0$ and $0$ otherwise. In addition, for the payoff matrix (\ref{mat}) we  have
\begin{equation}
\Delta_m   =   \max_{\alpha,\beta} \left \{ \mid \alpha-\beta \mid \right \} ,
\end{equation}
where $\alpha=a,b$ and $\beta=c,d$.
Finally, adding the probabilities given in eqs.~(\ref{item1}) to  (\ref{item10}) yields
\begin{eqnarray} \label{Pd1}
P_i(t+\delta t)   & = &   P_i(t)  - \frac{1}{M} \sum_{\alpha,\beta} P_i^\alpha (t) \frac{J_\beta}{M-1}  \frac{\beta - \alpha}{\Delta_m} \nonumber \\ 
& & + \frac{1}{M} \sum_{\alpha,\beta} P_i^\alpha (t) \frac{J_\beta}{M-1} \theta(\alpha-\beta)  \frac{\beta - \alpha}{\Delta_m} \nonumber \\
&  &  + \frac{1}{M} \sum_{\alpha,\beta} Q_i^\beta(t)  \frac{I_\alpha}{M-1} \theta(\alpha-\beta)  \frac{ \alpha-\beta}{\Delta_m} . \nonumber \\
& &
\end{eqnarray}
Since  $P_i (t+ \delta t) - P_i (t)$ must be proportional to $\delta t$, we must set $\delta t = 1/M$, as expected. 
To proceed, note that $P_i^\alpha$ and $Q_i^\beta$ are the same for all individuals, i.e., there is nothing in the model formulation that distinguishes the individuals a priori. Thus, we write  $P_i^\alpha (t) = P^\alpha (t) $ and $Q_i^\beta (t) = Q^\beta (t) $. In the limit $M\to \infty$
we can use the law of the large numbers \cite{Feller_1968} to write $P^\alpha (t) = I_\alpha/M$ and $Q^\beta (t)  = J_\beta/M$, so that
the third and the fourth terms on the right-hand side of eq.~(\ref{Pd1}) cancel each other out. With these observations, we rewrite eq.~(\ref{Pd1})  as 
\begin{eqnarray}\label{eq15}
 \frac{dP}{dt}&  = &  - \frac{1} {\Delta_m} \sum_{\alpha,\beta} P^\alpha (t) Q^\beta (t)  \left ( \beta - \alpha \right )  \nonumber \\
& = &   - \frac{1} {\Delta_m} \left [ P(t) \sum_\beta Q^\beta(t) \beta - Q(t)  \sum_\alpha P^\alpha(t) \alpha \right ] . \nonumber \\
& &
\end{eqnarray}
To complete the proof, we note that 
\begin{equation}
\pi^A = \frac{1}{P(t)} \sum_{\alpha} P^\alpha (t) \alpha 
\end{equation}
is the expected payoff of a player using strategy $A$ (or the average payoff conditional on using strategy $A$).  A similar expression holds for the expected payoff of a player using strategy $B$,
\begin{equation}
\pi^B = \frac{1}{Q(t)} \sum_{\beta} Q^\beta (t) \beta .
\end{equation}
Therefore
\begin{eqnarray}\label{fin}
\frac{dP}{dt}  &  = &   \frac{1} {\Delta_m}  P (t) Q(t)  \left ( \pi^A - \pi^B  \right )  \nonumber \\
& = &   \frac{1} {\Delta_m}  P (t) \left [ 1 - P(t) \right ] \left ( \pi^A - \pi^B  \right ) ,
\end{eqnarray}
which, apart from a trivial time rescaling, is the replicator equation for a two-strategy game \cite{Hofbauer_1998,Nowak_2006}. 
It is clear from Eq. (\ref{eq15}) that if the payoff difference $\beta - \alpha$ were replaced by some nonlinear function   $g(\beta - \alpha)$, such as the Fermi function, then the replicator equation could not be recovered.


\section{$N$-player games}

Classical examples of $N$-player games are the $N$-player snowdrift game \cite{Zheng_2007,Pacheco_2009}, which is a generalization of the hawk-dove game, and the $N$-player prisoner's dilemma game \cite{Hannelore_2010,Sigmund_2010} (see \cite{Archetti_2012} for a review).  In a typical scenario of an $N$-player game, the instantaneous payoff of individual $i$ is obtained by randomly selecting $N-1$ other individuals in the population without replacement. The payoff of individual $i$ depends on its strategy ($A$ or $B$) and the composition of the play group, i.e., how many individuals play strategies $A$ and $B$ in the play group.  More concretely, let us assume that the possible payoffs are $\alpha = a_1, \ldots, a_K$ for individuals using strategy $A$,  and  $\beta = b_1, \ldots, b_L$ for individuals using strategy $B$. 
As in the  $2$-player games, this procedure is done for the $M$ individuals so that  at time $t$ the population is completely described by the number of individuals  playing strategy $A$ with payoff $\alpha$ ($I_\alpha$) and the number of individuals  playing strategy $B$ with payoff $\beta$ ($J_\beta$) where $\alpha$ and $\beta$ take all the possible payoff values for each strategy. With this notation, the derivation of the differential equation for $P(t)$ is identical to that outlined before for the $2$-person games, resulting in the replicator equation (\ref{fin}).

\section{Conclusion}

Evolutionary game theory, introduced in the early 1970s to explain the ritualistic nature of many animal contests \cite{Maynard_1973} 
has opened up an entire field of research with implications far beyond biology \cite{Traulsen_2023}.  The finding that the replicator equation, called the equation of life for its central role in evolutionary game theory \cite{Nowak_2006}, describes the infinite population limit of the eminently social process of copying or imitating  peers \cite{Traulsen_2005} has contributed greatly to this success.  Here we provide an alternative  elementary (and admittedly non-rigorous) proof of the link between the imitation dynamics in the large population limit and the replicator equation for the  scenario where players compare their instantaneous (not average) payoffs and only more successful individuals can be imitated. Our proof can be easily generalized to the case where players can use more than two strategies. 

\bigskip

\acknowledgments

This research  was partially supported by  Conselho Nacional de Desenvolvimento Ci\-en\-t\'{\i}\-fi\-co e Tecnol\'ogico -- Brasil (CNPq) -- grant number 305620/2021-5.

\end{document}